\documentclass[twocolumn]{aastex631}

\begin{document}

\title{A Multi-messenger Search for Ultra-high-energy Gamma Rays in Coincidence with Neutrinos }

\author{R.~Alfaro}
\affiliation{Instituto de F\'{i}sica, Universidad Nacional Autónoma de México, Ciudad de Mexico, Mexico }
\author{C.~Alvarez}
\affiliation{Universidad Autónoma de Chiapas, Tuxtla Gutiérrez, Chiapas, México}
\author{A.~Andr\'{e}s}
\affiliation{Instituto de Astronom\'{i}a, Universidad Nacional Autónoma de México, Ciudad de Mexico, Mexico }
\author{E.~Anita-Rangel}
\affiliation{Instituto de Astronom\'{i}a, Universidad Nacional Autónoma de México, Ciudad de Mexico, Mexico }
\author[0000-0002-0595-9267]{M.~Araya}
\affiliation{Universidad de Costa Rica, San José 2060, Costa Rica}
\author{J.C.~Arteaga-Velázquez}
\affiliation{Universidad Michoacana de San Nicolás de Hidalgo, Morelia, Mexico }
\author[0000-0002-4020-4142]{D.~Avila Rojas}
\affiliation{Instituto de Astronom\'{i}a, Universidad Nacional Autónoma de México, Ciudad de Mexico, Mexico }
    \author[0000-0002-2084-5049]{H.A.~Ayala Solares}
\affiliation{Temple University, Department of Physics, 1925 N. 12th Street, Philadelphia, PA 19122, USA }
    \author[0000-0002-5529-6780]{R.~Babu}
\affiliation{Department of Physics and Astronomy, Michigan State University, East Lansing, MI, USA }
\author{P.~Bangale}
\affiliation{Temple University, Department of Physics, 1925 N. 12th Street, Philadelphia, PA 19122, USA }
% ...




\begin{abstract}

The last five years have shown us that ultra-high-energy (UHE; $>$100 TeV) gamma-ray sources are ubiquitous, but the nature of these sources remain highly uncertain. UHE gamma rays can be produced via either leptonic (Inverse compton) or hadronic (pion decay) emission mechanisms. To decisively determine the emission mechanisms, multimessenger searches are essential. Neutrinos are of particular interest as they are only created via hadronic channels. In this work, we describe a metric to select high-quality UHE events from the High Altitude Water Cherenkov (HAWC) Observatory. We use this metric to search for correlations between HAWC archival data and IceCube public neutrino alerts. 24 spatial coincidences are found, which is higher than the number of events expected by random chance. Therefore, we conclude that there are likely associations between HAWC gamma rays and IceCube neutrinos, but the angular resolutions of the two instruments prevent us from conclusively making any definitive associations between the coincidences and specific astrophysical sources. More sensitive detectors are needed.

\end{abstract}

\keywords{Gamma-ray astronomy (628), Gamma-ray sources (633), Gamma-ray transient sources (1853), Neutrino astronomy (1100), High energy astrophysics (739), Particle astrophysics (96)}


\section{Introduction} \label{sec:intro}

Multi-messenger astrophysics combines information using different astrophysical messengers, such as electromagnetic observations, neutrinos, and gravitational waves, to determine the most complete picture of a source. After the observation of both gamma rays and neutrinos from Supernova 1987A~\citep{1987PhRvL..58.1490H, 1988Natur.332..516L}, there was a decades-long gap before more multi-messenger events were observed. The current dawn of multi-messenger astrophysics began within the last decade. Notable events include the joint detection of gravitational wave GW170817 coincident with gamma-ray burst 170817A ~\citep{2017ApJ...848L..13A}, the observation of a flaring gamma-ray blazar, TXS 0506+056, coincident with neutrino IceCube-170922A ~\citep{2018Sci...361.1378I}, and the detection of the Galactic plane in both neutrinos~\citep{2023Sci...380.1338I} and electromagnetic wavelengths.

The first detection of an ultra-high-energy (UHE; $>$ 100 TeV) source was of the Crab Nebula, by the Tibet air shower array~\citep{PhysRevLett.123.051101}. Since then, observations from both the High Altitude Water Cherenkov (HAWC) Observatory and the Large High Altitude Air Shower Observatory (LHAASO) have shown that gamma-ray sources are ubiquitous at these energies~\citep{2020PhRvL.124b1102A,2021ApJ...911L..27A,2021Natur.594...33C}. These gamma rays are particularly interesting for multi-messenger astronomy.  These gamma rays are expected to be Galactic in origin, as extragalactic gamma rays of this energy are attenuated before reaching the Earth's surface, and could be associated with hadronic PeVatrons and provide clues to the origin of Galactic cosmic rays. However, emission mechanisms remain unclear, as both leptonic and hadronic processes can create gamma rays above this energy, and relatively low statistics make it difficult to get a clear picture of the spectrum above this threshold. Many observations begin to have poor statistics near the beginning of the regime where leptonic and hadronic models are expected to diverge (see \cite{2022ApJ...928..116A} for an example). For this reason, joint searches with neutrinos are advantageous, as all neutrinos are hadronic in origin. The superior angular resolution and relatively high significance of gamma ray sources can be combined with the unambiguous origin of neutrinos to create a joint analysis that is greater than the sum of its parts. Note that many experiments have instituted real-time multi-messenger searches, for example, ~\cite{2024JCAP...09..042A}, but none are optimized exclusively on UHE events. ~\cite{2024Univ...10..326S} and \cite{AARTSEN201730}provide an overview of IceCube real-time alert programs. The authors note that this alert pipeline was used to discover the neutrino-emitting blazar TXS 0506+056~\citep{2018Sci...361.1378I}, but no other neutrino sources have been conclusively identified via the IceCube alerts.  

UHE gamma rays overlap well with the energy range of the IceCube Neutrino Observatory. The dominant production mechanism here is proton-proton interactions induced when protons interact with nearby material, such as a molecular cloud. This creates both charged and neutral pions, with approximately a 2:1 ratio. The charged pions then decay into neutrinos, each of which have an energy that is roughly one quarter of the charged pion energy. The neutral pions decay into gamma ray that have roughly half the energy of the pion. Therefore, there is a factor of two difference in the energy of a gamma ray and the energy of a neutrino created via the same hadronic mechanisms~\citep{2014PhRvD..90b3010A,Kelner:2008ke}. 

In this work, we present a real-time UHE alert system for the HAWC Observatory. This alert system selects high-quality events that are likely to be UHE gamma rays with high confidence (as opposed to a mis-reconstructed cosmic ray or mis-reconstructed lower-energy gamma ray). We decide a priori that we wish to send approximately two alerts per month, which we believe to be a reasonable rate as to not overwhelm HAWC's partners with requests for follow-up observations. We test the performance of the system on archival data, including quantifying the false positive rate.  This alert system is mainly intended to be used in multi-messenger searches. However, for bright enough steady sources it may be able to be used in a standalone manner to detect Galactic transients. For a review of high-energy Galactic transients, see ~\cite{2024Univ...10..163C}. 

We then look for spatial coincidences between HAWC events found with our metric and public neutrino alerts from the IceCube Neutrino Observatory. We use nearly 9 years of HAWC data and 13 years of IceCube alerts for this study.

The paper is organized as follows: Section \ref{sec:obs} describes the HAWC and IceCube observatories. Section \ref{sec:stat} describes the HAWC alert metric. Section \ref{sec:archival} describes the analysis with archival HAWC and IceCube data. Conclusions are presented in Section \ref{sec:conclusions}.

\section{Observatory Information}\label{sec:obs}

HAWC is a TeV gamma-ray observatory located at an altitude of 4100 meters in the state of Puebla, Mexico. The experiment consists of an array of 300 water Cherenkov tanks, each outfitted with 4 PMTs. There are also several smaller outriggers with one PMT each. The outriggers are not considered in this analysis.

HAWC has a large field-of-view and high duty cycle, observing $>$ 70$\%$ of the sky daily with uptime of approximately 95$\%$. This makes the experiment a good choice for all-sky analyses and searches for transients.  HAWC is sensitive to gamma rays from approximately 300 GeV to several hundred TeV. A fast-running reconstruction runs online at the HAWC site and is used to search for transients in real-time, with latency on the order of seconds.  The raw data is then transferred off-site and reconstructed again. The main differences between the online and offline reconstruction include a better determination of bad or inoperable PMTs. For more information about the HAWC detector, see \cite{2023NIMPA105268253A}. 

IceCube is a neutrino observatory located at the South Pole. The observatory consists of an array of 5160 digital optical modules arranged on 86 vertical strings encased in the Antarctic ice, each of which is equipped with 60 digitial optical modules (DOMs). Each DOM is a glass pressure vessel containing a 10-inch PMT. Like HAWC, these PMTs also detect Cherenkov light, although instead of water Cherenkov tanks, the detection medium is the Antarctic ice. IceCube also has close to 100 $\%$ uptime. IceCube is sensitive to neutrinos in the GeV to PeV energy ranges, with the GeV sensitivity coming from a low-energy subarray called DeepCore. IceCube has observed the full sky in its fully constructed 86-string configuration since 2010, with additional data from the partially constructed array starting in 2008. For more information about the IceCube detector, see \cite{2017JInst..12P3012A}.

\section{The HAWC UHE alert metric}\label{sec:stat}

The event rate of UHE gamma rays, while relatively small, is still too large to notify all of HAWC's partners of each event for potential follow-up. Above 177 TeV, approximately 1.3 events per day pass gamma/hadron separation cuts. Not all of these detected events are high quality. Some may have a poorly known energy estimate or reconstruction parameters that are more indicative of a cosmic ray.

To select single high-quality events that should be investigated further, we develop a metric that takes into account specific reconstruction parameters to select events that are both high energy and gamma-like. This metric has a range between 0 and 50, with 50 being the highest possible score. The algorithm to calculate the metric only reports integer values.

\subsection{Alert metric criteria}\label{sec:criteria}

The inputs to the metric can roughly be divided into three main categories: gamma-ness, energy, and location. The score is determined based on the values of various reconstruction variables related to each of those categories. These are described in more detail below.  Table \ref{tab:statistic} shows the maximum score possible for each of the criteria, with elaborations on each parameter below. 

\begin{table*}
    \centering
    \begin{tabular}{|c|c|c|}
         \hline
         Parameter & Value  & Points Awarded \\
         \hline
         ChiSq & $\leq$ ChiSq$_{max}$  & 10 \\
         ChiSq & $\leq$ ChiSq$_{max}$ - 0.05 & 5  \\
         ChiSq & $\leq$ ChiSq$_{max}$ - 0.1 & 5 \\
         Location & $\leq$ 0.5 deg. from nearest TeVCat source & 20 \\
         Location & $0^{\circ} < l < 90^{\circ}$ and $|b| < 3^{\circ}$ and $\geq$ 0.5 deg. from TeVCat source & 10 \\
         Location & $0^{\circ} < l < 90^{\circ}$ and $|b| < 1^{\circ} $ and $\geq$ 0.5 deg. from TeVCat source & 10  \\
         Energy & E $>$ 56 TeV after subtracting energy resolution  & 2 \\
         Energy & E $>$ 100 TeV after subtracting energy resolution & 2 \\
         Energy & E $>$ 177 TeV after subtracting energy resolution & 2 \\
         Energy & E $>$ 316 TeV after subtracting energy resolution & 2\\
         Energy agreement & GP and NN within 2 sigma & 1 \\
         Energy agreement & GP and NN within 1 sigma & 1 \\
         \hline
    \end{tabular}
        \caption{Each parameter that contributes to the alert metric and the number of points awarded for meeting that criteria. The maximum score is 50. For the chi square, the minimum cut is both zenith- and time-dependent (see Section \ref{sec:syst} for a definition of ChiSq$_{max}$). $l$ and $b$ are Galactic coordinates. Note that the maximum score that can be awarded based on the location is 20. An event cannot receive double points for both being associated with a TeVCat source and within a degree of the Galactic plane. Also note that the energy scores are additive. For example, an event with an energy of 58 TeV would receive 2 points, while an event with an energy of 101 TeV would receive 4 points (2 from the $>$56 TeV criteria and 2 from the $>$ 100 TeV criteria. If the energy is above 500 TeV, the value of the metric is artificially set to 0 as the energy estimate is unreliable above these energies.}
    \label{tab:statistic}
\end{table*}

\textbf{Gamma-ness}: HAWC has a very large cosmic ray background. The main determinant of whether a detected particle is gamma-like or a cosmic ray is the chi square of the fit to the lateral distribution function (LDF). This is the main gamma/hadron separation parameter currently used in HAWC at high energies~\citep{2024ApJ...972..144A}.  A lower chi square value indicates a higher likelihood that the event is a gamma ray. This is because gamma ray LDFs do not have muons that land far from the air shower core, while cosmic ray LDFs do. See \cite{2019ApJ...881..134A} for more information. In this analysis, the smaller the chi square value, the higher score the event receives in the metric calculation.  The exact chi square used is zenith-angle dependent to account for declination-dependent exposure effects.  Due to a data quality issue that has caused the chi square to drift over time, the chi square cut is also time-dependent. This is discussed further in the systematics section (Section \ref{sec:syst}).

\textbf{Energy}: Higher energy events have a higher metric value by default (i.e. a 200 TeV gamma ray will have a higher value than a 60 TeV gamma ray). HAWC has two energy estimators in use. The first, known as the ``ground parameter" (GP), uses the charge in the PMTs a specified distance from the shower core, along with the zenith angle of the event, to estimate the energy. The second is a neural network (NN) which uses the amount of charge in annuli around the shower core, among other parameters, as the inputs.  In order to receive the maximum score, the ground parameter and neural network energy estimates must be close together (within the energy resolution of the energy estimators). Both estimators are discussed in-depth in ~\cite{2019ApJ...881..134A}. 

The energy resolution was taken from \cite{2019ApJ...881..134A}; the formula to determine whether any points are awarded is

\begin{equation}
    (E_{gp} - E_{nn})/E_{res, gp} < \sigma_i,
\end{equation}
where $E_{gp}$ is the ground parameter energy, $E_{nn}$ is the neural network energy, $E_{res, gp}$ is the energy resolution of the ground parameter energy at that energy, and $\sigma$ is the sigma level. $i$ can be 1 or 2, corresponding to the 1$\sigma$ and 2$\sigma$ energy agreement thresholds shown in the last two rows of Table \ref{tab:statistic}.

Note that any event with either a GP or NN energy estimate above 500 TeV will have a metric of 0. This is because HAWC photons are not simulated above this energy threshold, and these energy estimates are inherently unreliable. 

\textbf{Location}: Events that are near known sources are more likely to be gamma rays. Events receive the full amount of location points if they are either within 0.5 degrees of a known TeVCat\footnote{http://tevcat.uchicago.edu/} source~\citep{2008ICRC....3.1341W} or within 1 degree of the inner Galactic plane (defined as 0 $<$ $l$ $<$ 90 degrees).  Events receive a lower score if they are within 3 degrees of the inner Galactic plane, but not within 1 degree of the Galactic plane or near a TeVCat source. This ensures that events that could possibly be attributed to diffuse emission in the Galactic plane contribute to the metric. Recent papers, including one from the Tibet AS$\gamma$ collaboration, have shown that a non-negligible number of UHE events in the Galactic disk can be attributed to diffuse emission~\citep{2021PhRvL.126n1101A}. 

The final score is the sum of all the parameters seen in Table \ref{tab:statistic} (i.e. the gamma-like + location + energy + energy agreement scores, which are 40$\%$, 40$\%$, 16$\%$, and 4$\%$ of the total, respectively).

\subsection{Alert threshold}

We decide a priori that we wish to send approximately 2 alerts per month, which we believe to be a reasonable rate for HAWC's partners to have the capacity to do follow-up observations for. Looking at 2800 days archival data from June 2015 to January 2024, we select events that pass both HAWC's standard data quality cuts (i.e. the event must have a successfully reconstructed core location and angle) as well as loose energy cuts of $E_{gp} > $ 56 TeV and $E_{nn} > $ 56 TeV. 73,389 events pass this criteria. 

\begin{figure}
    \centering
    \includegraphics[width=0.98\linewidth]{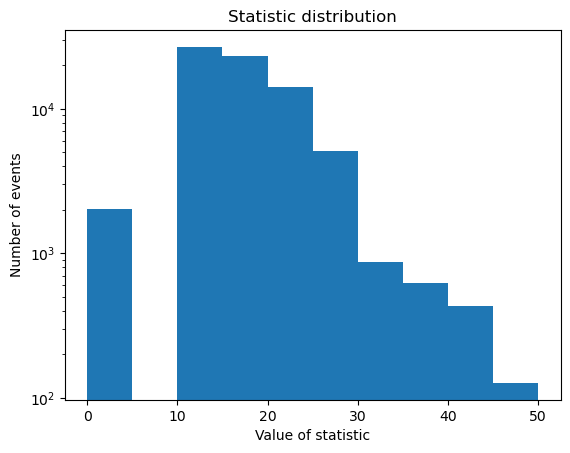}
    \caption{The distribution of the metric value for all UHE events passing the standard analysis cuts and a loose energy cut. The spike at zero is due to events that have either a GP or NN energy estimate above 500 TeV, where simulations are not available for comparison to data.}
    \label{fig:hist}
\end{figure}

We calculate the metric for each of these events. A histogram of this metric is shown in Figure \ref{fig:hist}. Using this histogram, an alert threshold of a score $\geq$ 45 is chosen to coincide with the previously chosen alert rate. 127 events have a metric $\geq$ 45, which translates to one alert approximately every 22.04 days. 

\subsection{False positive rate}

\begin{figure}
    \centering
    \includegraphics[width=0.98\linewidth]{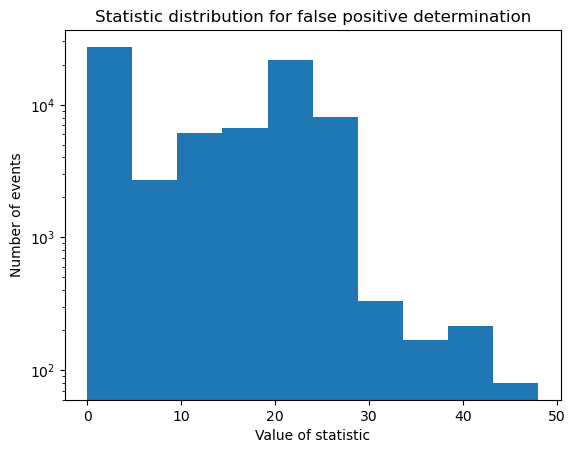}
    \caption{The metric calculated when each variable that is used in the calculation is randomly drawn according to that distribution. This is used in the false positive determination. }
    \label{fig:falsePos}
\end{figure}

We now quantify the false positive rate. In this context, a ``false positive" is either a mis-reconstructed lower-energy gamma ray, or a cosmic ray misclassified as a gamma ray. To determine this, we create histograms of the distribution of every reconstruction parameter that is used to quantify the alerts: the chi square, right ascension (RA), declination (dec), GP energy, NN energy, the zenith angle of the event, and the time of the event. We then sample from these distributions randomly 73,389 times (equal to the number of events that passed the loose cut string when we determined the threshold) and compute the alert metric for these randomly drawn numbers. We then repeat this exercise 1000 times, so that we have 1000 trials. This distribution for one of these trials can be seen in Figure \ref{fig:falsePos}. 

From the 1000 trials, a mean value of 26.257 events from the 73389 randomly drawn in each trial have a metric above the alert threshold of 45. The standard deviation is 5.141. This means that 26.257 $\pm$ 5.141/127 archival alerts (0.207 $\pm$ 0.040) are likely false positives, corresponding to a false alert every 106.47 $\pm$ 20.56 days or $\sim$3.43 $\pm$ 0.66 per year. 

\subsection{Systematic uncertainties}\label{sec:syst}

We quantify the effect of a few systematic uncertainties on the analysis. 

\textbf{Chi square drift over time}: As mentioned in Section \ref{sec:criteria}, chi square distribution arising from the fit to the lateral distribution function began changing in 2018. The cause of this is presently unknown, but it is likely that this data quality issue arises from problems from inadequately quantifying bad or aging PMTs, which subsequently prevents those PMTs from being removed from the reconstruction.

All of the photons used in this analysis have extremely low chi square values (i.e. those on the left-hand side of the chi square distribution, in the tails). To quantify the change in the average chi square of these events, we take one day of data from September 2018 and select the events with an energy above 56 TeV.  We then take the lowest ten chi square values from that run and average them.  We repeat this exercise for every 12 months after. 

\begin{figure}
    \centering
    \includegraphics[width=0.98\linewidth]{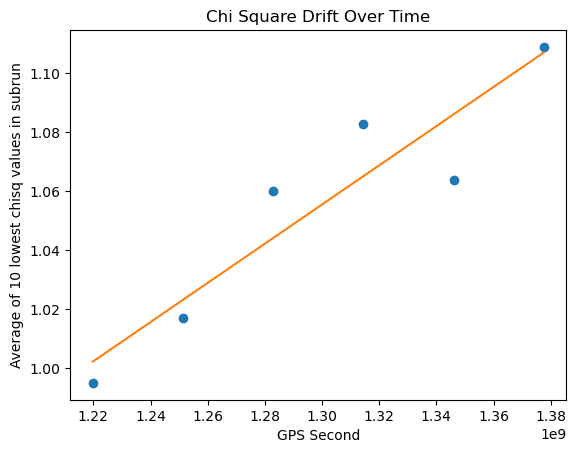}
    \caption{The average of the ten lowest chi squares in a run from September 2018, and then every 12 months afterward. The x-axis is reported in GPS Seconds, or seconds since January 1, 1980. September 2018 corresponds to approximately 1.22e9 GPS Seconds. This is used to quantify how much the left-hand (extremely gamma-like) side of the distribution is shifting in time. This effect is treated as linear in the analysis. Uncertainties are neglible. }
    \label{fig:chiDrift}
\end{figure}

Figure \ref{fig:chiDrift} shows these values plotted as a function of time. It can be seen that this effect is roughly linear (orange line). 

To account for this effect in the analysis, we implement a time-dependent chi square cut. Before mid-2018, the maximum chi square allowed in the analysis is

\begin{equation}
    1.0 + 0.41683\theta^4,
\end{equation}
where $\theta$ is the zenith angle of the event in radians. The zenith angle parameterization is the standard zenith-angle parameterization for gamma/hadron separation cuts used in the pass 5 HAWC data reconstruction~\citep{2024ApJ...972..144A}. 

After mid-2018, the maximum chi square allowed in the analysis is 

\begin{equation}
    (6.6465 \times 10^{-10})s + 0.19149 + 0.41683\theta^4,
\end{equation}
where $s$ is the GPS second of the event.  The first two terms here come from the best-fit line to the change in the chi square over time (the orange line in Figure \ref{fig:chiDrift}).

\textbf{Chi square online vs. offline}: The maximum chi square cut was determined on archival data. We check to make sure that the distribution is not significantly different in HAWC's online reconstruction, where the alerts are currently running. The PMT hit selection is slightly different online. In this environment, which is a scaled back, faster version of the offline reconstruction, not all of the diagnostics to identify bad PMTs are run. See ~\cite{2018NIMPA.888..138A} for more information on the HAWC online processing system. 

\begin{figure}
    \centering
    \includegraphics[width=1\linewidth]{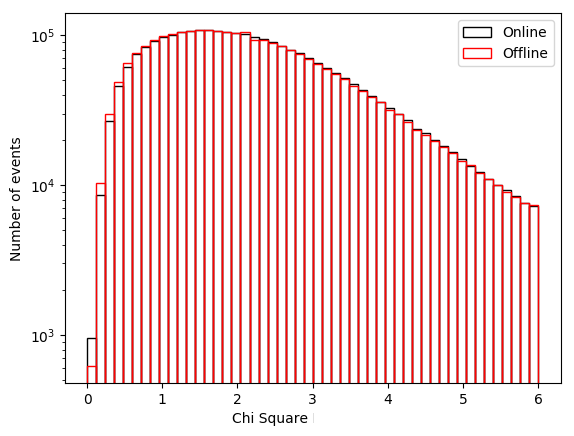}
    \caption{The chi square distribution for the HAWC offline reconstruction (red) vs. the online reconstruction (black). }
    \label{fig:onlineDiff}
\end{figure}

Figure \ref{fig:onlineDiff} shows the online reconstruction vs. offline reconstruction distribution of the chi square. Note that there are slight differences in the left-hand side of the plot. 

We then check to see what effect this has on the analysis.  We select events passing the standard analysis cuts, and count how many events pass the chi square cut. For this selection of data (which is from January 2024), the number of events passing the offline analysis data is 3.2$\%$ higher. 

The threshold set using archival data corresponds to one event approximately every 22.04 days. The expected rate increase online translates to a new event rate of one event every 21.37 days, a negligible change. We conclude that slight differences in the online vs. offline analysis do not affect the calculations presented here. 

\section{Archival search with IceCube alerts}\label{sec:archival}

The HAWC alert system has been running online continuously since summer 2024. When an alert is detected, an XML file with relevant information is created. This XML file uses the VOEvent schema\footnote{\url{https://www.ivoa.net/documents/VOEvent/}}. The latency between HAWC detecting an event and writing the XML file is less than 10 seconds.

To show the potential capabilities of this system, we perform a search using archival data along with IceCube public alerts. HAWC and IceCube have similar energy ranges and a sizable overlap in their field-of-view. A joint detection would help determine the origin of the emission of HAWC's previously observed UHE sources.

We use both IceCube bronze and gold alerts (which respectively have a 30$\%$ and 50$\%$ percent chance of having astrophysical origin), consisting of the events discussed in the IceCube alert catalog~\citep{2023ApJS..269...25A} along with additional alerts that have been distributed since the publication of that paper\footnote{Taken from here: \url{https://gcn.gsfc.nasa.gov/amon_icecube_gold_bronze_events.html} }. All of these alerts are track alerts, i.e. cascade events are not included. The time period of the IceCube alerts covers May 2011 through April 2024. There are 357 alerts during this time period. Note that this is a different time period than the HAWC alerts, which cover June 2015 through January 2024. However, since we are interested in spatial coincidences, we do not trim the data for either experiment as to not exclude any potential spatial coincidences. 

\subsection{Method and results}

We take the locations of the IceCube alerts along with the containment radius. We then flag every HAWC alert that is within the containment radius of an IceCube alert. Therefore, one HAWC alert can potentially be associated with multiple IceCube alerts, and vice versa. 

24 spatial coincidences are found. The smallest difference in time is approximately one month. with an IceCube alert arriving on November 20, 2020 and a HAWC UHE photon being detected on December 18, 2020.   A list of these coincidences, with location information, can be found in Table \ref{tab:alerts}.  Table \ref{tab:energy} contains the timing and energy information for each alert. 7 distinct IceCube alerts are associated with a HAWC alert.  All seven of these alerts are bronze events. 

\begin{table*}
\begin{tabular}{llllllll}
Coincidence & & & & & HAWC  \\ Number & IC RA  & IC Dec & HAWC RA & HAWC Dec & metric & $d$ & $c$  \\
1            & 307.66 & 40.72  & 307.25  & 37.89    & 50             & 2.85                              & 4.66                  \\
2            & 86.99  & 48.83  & 95.35   & 38.63    & 48             & 11.84                             & 12.63                 \\
3            & 307.66 & 40.72  & 305.31  & 36.84    & 48             & 4.29                              & 4.66                  \\
4            & 307.66 & 40.72  & 306.98  & 41.02    & 50             & 0.6                               & 4.66                  \\
5            & 307.66 & 40.72  & 304.95  & 36.67    & 46             & 4.57                              & 4.66                  \\
6            & 286.92 & 6.43   & 286.87  & 6.32     & 46             & 0.12                              & 0.94                  \\
7            & 287.18 & 5.53   & 286.87  & 6.32     & 46             & 0.85                              & 0.95                  \\
8            & 286.92 & 6.43   & 286.95  & 6.21     & 45             & 0.22                              & 0.94                  \\
9            & 287.18 & 5.53   & 286.95  & 6.21     & 45             & 0.72                              & 0.95                  \\
10           & 286.92 & 6.43   & 286.87  & 6.26     & 46             & 0.18                              & 0.94                  \\
11           & 287.18 & 5.53   & 286.87  & 6.26     & 46             & 0.79                              & 0.95                  \\
12           & 286.92 & 6.43   & 287.08  & 6.3      & 46             & 0.21                              & 0.94                  \\
13           & 287.18 & 5.53   & 287.08  & 6.3      & 46             & 0.78                              & 0.95                  \\
14           & 307.66 & 40.72  & 305.67  & 38.12    & 46             & 3.02                              & 4.66                  \\
15           & 265.52 & 7.33   & 265.03  & 9.83     & 45             & 2.55                              & 2.92                  \\
16           & 287.18 & 5.53   & 286.29  & 5.37     & 46             & 0.9                               & 0.95                  \\
17           & 286.92 & 6.43   & 286.82  & 6.08     & 46             & 0.36                              & 0.94                  \\
18           & 287.18 & 5.53   & 286.82  & 6.08     & 46             & 0.66                              & 0.95                  \\
19           & 287.18 & 5.53   & 286.65  & 5.23     & 46             & 0.61                              & 0.95                  \\
20           & 307.66 & 40.72  & 309.78  & 41.21    & 46             & 1.67                              & 4.66                  \\
21           & 286.92 & 6.43   & 286.67  & 5.92     & 46             & 0.57                              & 0.94                  \\
22           & 287.18 & 5.53   & 286.67  & 5.92     & 46             & 0.64                              & 0.95                  \\
23           & 324.58 & 51.74  & 316.2   & 47.56    & 46             & 6.84                              & 8.25                  \\
24           & 280.46 & -1.9   & 281.53  & -2.89    & 50             & 1.46                              & 1.49                          
\end{tabular}
\caption{Information about each of the spatial coincidences between IceCube and HAWC.  All locations, distances, and the containment radius are in degrees.  $d$ is the distance between the HAWC and IceCube locations. $c$ is the reported IceCube containment radius. All of the HAWC events are correlated with IceCube Bronze alerts. No correlations with Gold alerts were found.}
\label{tab:alerts}
\end{table*}

\begin{table*}[]
\begin{tabular}{llllll}
Coincidence Number & IceCube time   & HAWC time      & $\Delta$t (days) & IceCube energy (TeV) & HAWC energy (TeV) \\
1                  & 11/20/20 9:44  & 7/15/15 6:41   & 1955.13       & 154                  & 420               \\
2                  & 8/12/16 16:25  & 4/4/16 19:43   & 129.86        & 160                  & 421               \\
3                  & 11/20/20 9:44  & 8/17/16 3:53   & 1556.24       & 154                  & 252               \\
4                  & 11/20/20 9:44  & 1/2/17 20:08   & 1417.57       & 154                  & 437               \\
5                  & 11/20/20 9:44  & 3/18/17 15:41  & 1342.75       & 154                  & 162               \\
6                  & 1/19/15 8:50   & 10/19/17 0:04  & -1003.63      & 140                  & 143               \\
7                  & 3/13/18 16:17  & 10/19/17 0:04  & 145.68        & 160                  & 143               \\
8                  & 1/19/15 8:50   & 10/29/17 0:56  & -1013.67      & 140                  & 172               \\
9                  & 3/13/18 16:17  & 10/29/17 0:56  & 135.64        & 160                  & 172               \\
10                 & 1/19/15 8:50   & 12/11/17 19:07 & -1057.43      & 140                  & 210               \\
11                 & 3/13/18 16:17  & 12/11/17 19:07 & 91.88         & 160                  & 210               \\
12                 & 1/19/15 8:50   & 6/4/18 11:29   & -1232.11      & 140                  & 158               \\
13                 & 3/13/18 16:17  & 6/4/18 11:29   & -82.80        & 160                  & 158               \\
14                 & 11/20/20 9:44  & 4/11/19 9:24   & 589.01        & 154                  & 194               \\
15                 & 11/23/21 14:25 & 6/16/19 4:38   & 891.41        & 142                  & 189               \\
16                 & 3/13/18 16:17  & 7/10/19 5:47   & -483.56       & 160                  & 146               \\
17                 & 1/19/15 8:50   & 6/4/20 9:29    & -1963.03      & 140                  & 147               \\
18                 & 3/13/18 16:17  & 6/4/20 9:29    & -813.72       & 160                  & 147               \\
19                 & 3/13/18 16:17  & 9/6/20 2:52    & -907.44       & 160                  & 184               \\
20                 & 11/20/20 9:44  & 12/18/20 20:49 & -28.46        & 154                  & 182               \\
21                 & 1/19/15 8:50   & 6/21/22 7:46   & -2709.96      & 140                  & 177               \\
22                 & 3/13/18 16:17  & 6/21/22 7:46   & -1560.65      & 160                  & 177               \\
23                 & 11/20/18 22:38 & 8/22/22 8:59   & -1370.43      & 173                  & 143               \\
24                 & 1/25/13 15:48  & 9/20/22 1:21   & -3524.40      & 114                  & 442              
\end{tabular}
\caption{Information about the time and energy of each coincidence. The all reported times are in UTC. The `HAWC Energy' column is the average of HAWC's two energy estimators.}
\label{tab:energy}
\end{table*}

We also look at the nearest TeVCat source for each spatial coincidence. This information can be found in Table \ref{tab:tevcat}. Note that the nearest TeVCat source is calculated using the position of the HAWC photon. This choice is made because the HAWC detector has better angular resolution than IceCube. 

\begin{table*}[]
\centering
\begin{tabular}{lll}
Coincidence Number & Nearest TeVCat Source & Distance to TeVCat source (deg) \\
1                  & MilagroDiffuse    & 2.25                      \\
2                  & 3HWC J0621+382        & 0.42                      \\
3                  & VER J2019+368         & 0.46                      \\
4                  & MGRO J2031+41         & 0.35                      \\
5                  & VER J2019+368         & 0.17                      \\
6                  & MGRO J1908+06         & 0.12                      \\
7                  & MGRO J1908+06         & 0.12                      \\
8                  & MGRO J1908+06         & 0.06                      \\
9                  & MGRO J1908+06         & 0.06                      \\
10                 & MGRO J1908+06         & 0.11                      \\
11                 & MGRO J1908+06         & 0.11                      \\
12                 & LHAASO J1908+0621     & 0.06                      \\
13                 & LHAASO J1908+0621     & 0.06                      \\
14                 & MilagroDiffuse    & 0.68                      \\
15                 & 3HWC J1739+099        & 0.11                      \\
16                 & 2HWC J1902+048*       & 0.93                      \\
17                 & MGRO J1908+06         & 0.24                      \\
18                 & MGRO J1908+06         & 0.24                      \\
19                 & SS 433                & 1.02                      \\
20                 & LHAASO J2032+4102     & 1.74                      \\
21                 & MGRO J1908+06         & 0.46                      \\
22                 & MGRO J1908+06         & 0.46                      \\
23                 & RGB J2056+496         & 2.92                      \\
24                 & HESS J1846-029        & 0.11                     
\end{tabular}
\caption{Information about the nearest TeVCat source to each HAWC alert that has a spatial coincidence with an IceCube alert. The distance is based on the HAWC photon, not the IceCube alert location, as the HAWC photon is expected to have better localization. }
\label{tab:tevcat}
\end{table*}

12 of the 24 spatial coincidences are close to MGRO J1908+06. It is worth noting that MGRO J1908+06 is one of HAWC's highest-energy sources~\citep{2022ApJ...928..116A} and has long been thought of as a potential neutrino source, with it having among the highest $p$-values in an IceCube search for neutrino emission in the Galactic plane~\citep{2019EPJC...79..234A}. A detection of this source may be possible in neutrinos soon~\citep{2021PhRvD.103j3020N}, although depending on the cutoff, IceCube Gen2 may be required. Ten of these coincidences spatially correlated with MGRO J1908+06 are within 0.25 degrees, while the last two are 0.46 degrees away from the center of the source. Since MGRO J1908+06 has a Gaussian extent of 0.67 $\pm$ 0.03 degrees~\citep{2020PhRvL.124b1102A}, these can all be considered spatial coincidences with the source. 

The next most common region is the Cygnus region. Several TeVCat sources are located in this region: MilagroDiffuse, VER J2019+368, LHAASO J2032+4102 and MGRO J2031+41. There are six spatial coincidences from this region. Given our knowledge of the region, it is likely that many of these coincidences are actually associated with the Cygnus Cocoon, which is extremely extended and has been observed to be accelerating protons to at least a PeV~\citep{2021NatAs...5..465A,2024SciBu..69..449L}. However, some of these could come from the pulsar wind nebula (PWN) in the region~\cite{2024ApJ...975..198A}. It has been postulated that proton acceleration could be associated with PWN, which would imply neutrinos from these sources~\citep{2003A&A...402..827A}. There is also the possibility that these coincidences come from unresolved sources in the region.

In some cases, it appears that the HAWC gamma ray can be attributed to diffuse emission (either isotropic or Galactic in nature) or unresolved sources rather than a known TeVCat source. For example, the nearest TeVCat source to coincidence 23 is RGB J2056+496, which is a blazar observed in TeV by VERITAS~\citep{2017AIPC.1792e0001B}. However, the HAWC gamma ray is nearly 3 degrees away from this source, much larger than HAWC's angular resolution at these energies. Additionally, HAWC is not expected to detect UHE gamma rays from extragalactic sources, as they should be attenuated on their way to the Earth. It is also worth noting that the IceCube containment radius for this event is quite large (8.25 degrees). These factors together mean it is unlikely that the coincidence is associated with the blazar. 

\subsection{False coincidence rate calculation}

We investigate whether the number of spatial coincidences is above what is expected from background. We take the distributions of the HAWC RA and declination, the IceCube RA and declination, and the IceCube containment radius.  Drawing randomly from these distributions 1000 times, for 1000 trials and seeing how often the HAWC location is within the IceCube containment radius, we determine that the average number of random spatial coincidences is 0.002, with a standard deviation of 0.04. We interpret this to mean that we are seeing more spatial coincidences than we expect due to chance, although the angular resolutions of the two instruments and the present statistics prevent us from drawing any firm conclusions about which sources may be seeing coincidences. Note that this number does not take into account any declination-dependence in the IceCube containment radius.

We calculate the expected number of coincidences for the two regions where there are the greatest number of spatial coincidences: the MGRO J1908+06 region, and the Cygnus region. We take the distribution of right ascension and declination for the HAWC alerts (which are used for location as HAWC has been angular resolution than IceCube).  We pull a random (right ascension, declination) pair from this distribution and calculate the distance to MGRO J1908+06.  We then repeat this 1000 times, for 1000 trials.  Most trials have no alerts within 0.25 degrees of MGRO J1908+06, and none have more than one spatial coincidence.  The average number of coincidences is 0.007, with a standard deviation of 0.083. We conclude that it is statistically unlikely for all ten of the spatial coincidences to be close to MGRO J1908+06 by chance.

We repeat the same test to determine the likelihood of seeing so many coincidences in the Cygnus region.  We use a 6.0 degree circle centered at right ascension=307.17 degrees, declination 41.17 degrees. Once again, we find that none of the 1000 trials performed has more than one alert in this region. On average, there were 0.003 coincidences with a standard deviation of 0.055.

\section{Conclusions}\label{sec:conclusions}

In this work, we have described an alert system for UHE photons for HAWC. Given the low event rate for photons of this energy, it is designed to detect single events from steady sources. However, if known TeV sources begins flaring at UHE energies, the system would also create and send alerts for that scenario. This system is fully operational, and we perform a search for spatial coincidences with IceCube alerts. 24 coincidences are found; this is above the number expected due to random chance. Here, we discuss the implications of this result as well as avenues for further study.

All of the photons passing the HAWC alert threshold are expected to be Galactic in nature. This belief arises from a combination of physics constraints (UHE photons from extragalactic sources are expected to attenuate before reaching the Earth) and the event selection criteria used (more weight is given to known TeVCat sources, the majority of which are Galactic in nature, and photons coming from regions of the Galactic plane). 

While the majority of IceCube neutrinos are extragalactic in origin and the Galaxy has been described as a ``neutrino desert"~\citep{2024NatAs...8..241F}, possibly related to the lack of black hole activity at the center of our Galaxy, the results here show that some fraction of the IceCube alerts are Galactic in origin.

We would also like to note that the alert system was optimized to send alerts for known gamma ray sources, but could easily have the weighting modified to send alerts for diffuse photons instead. This is of interest as the Tibet AS$\gamma$ collaboration has detected diffuse emission between 100 TeV and 1 PeV, with none of the observed gamma rays above 398 TeV being coincident with known sources~\citep{2021PhRvL.126n1101A}. Note that the UHE diffuse gamma-ray emission is likely not entirely leptonic in origin. For example, the contribution from unresolved pulsars has been ruled to be subdominant~\citep{2024ApJ...975L...6K}.

A study of UHE diffuse emission would not necessarily need to be in real time, as this component of the astrophysical emission is not expected to be transient in nature. Such a study would likely be archival.

\section{acknowledgments}
We acknowledge the support from: the US National Science Foundation (NSF); the US Department of Energy Office of High-Energy Physics; the Laboratory Directed Research and Development (LDRD) program of Los Alamos National Laboratory; Consejo Nacional de Ciencia y Tecnolog\'{i}a (CONACyT), M\'{e}xico, grants LNC-2023-117, 271051, 232656, 260378, 179588, 254964, 258865, 243290, 132197, A1-S-46288, A1-S-22784, CF-2023-I-645, CBF2023-2024-1630, c\'{a}tedras 873, 1563, 341, 323, Red HAWC, M\'{e}xico; DGAPA-UNAM grants IG101323, IG100726, IN111716-3, IN111419, IA102019, IN106521, IN114924, IN110521 , IN102223; VIEP-BUAP; PIFI 2012, 2013, PROFOCIE 2014, 2015; the University of Wisconsin Alumni Research Foundation; the Institute of Geophysics, Planetary Physics, and Signatures at Los Alamos National Laboratory; Polish Science Centre grant, 2024/53/B/ST9/02671; Coordinaci\'{o}n de la Investigaci\'{o}n Cient\'{i}fica de la Universidad Michoacana; Royal Society - Newton Advanced Fellowship 180385; Gobierno de España and European Union-NextGenerationEU, grant CNS2023- 144099; The Program Management Unit for Human Resources \& Institutional Development, Research and Innovation, NXPO (grant number B16F630069); Coordinaci\'{o}n General Acad\'{e}mica e Innovaci\'{o}n (CGAI-UdeG), PRODEP-SEP UDG-CA-499; Institute of Cosmic Ray Research (ICRR), University of Tokyo. H.M. acknowledges support under grant number CBF2023-2024-1630. H.F. acknowledges support by NASA under award number 80GSFC21M0002. C.R. acknowledges support from National Research Foundation of Korea (RS-2023-00280210). We also acknowledge the significant contributions over many years of Stefan Westerhoff, Gaurang Yodh and Arnulfo Zepeda Dom\'inguez, all deceased members of the HAWC collaboration. Thanks to Scott Delay, Luciano D\'{i}az and Eduardo Murrieta for technical support.

%

\vspace{5mm}
\facilities{HAWC, IceCube}





\bibliography{sample631}{}
\bibliographystyle{aasjournal}



\end{document}